\documentclass[conference]{IEEEtran}
\usepackage{cite}
\usepackage{amsmath,amssymb}
\usepackage{algorithm}
\usepackage{algpseudocode}
\usepackage{array}
\usepackage{booktabs}
\usepackage{graphicx}
\usepackage{microtype}
\usepackage{tabularx}
\usepackage{ragged2e}

\newcolumntype{L}[1]{>{\RaggedRight\arraybackslash\hsize=#1\hsize\linewidth=\hsize}X}
\newcommand{\grouphead}[1]{\multicolumn{5}{@{}l@{}}{\itshape #1}\\[1pt]}
\usepackage{dblfloatfix}

\usepackage[hidelinks]{hyperref}
\usepackage{xcolor}
\usepackage{xspace}
\newcommand{\pred}[1]{\mathsf{#1}}
\newcommand{\xkernel}{\textsc{Fabrica}\xspace}
\newcommand{\xbench}{\textsc{Fabrica-Bench}\xspace}
\newcommand{\fastzero}{\texttt{fast\_0}}
\newcolumntype{Y}{>{\raggedright\arraybackslash}X}

\begin{document}

\title{\xkernel: Agentic CUDA-to-CSL Translation and Optimization for
Wafer-Scale Systems}

\author{
\IEEEauthorblockN{Yuebo Luo\IEEEauthorrefmark{1},
Eliu Huerta\IEEEauthorrefmark{2}\IEEEauthorrefmark{3},
Venkatram Vishwanath\IEEEauthorrefmark{2},
Caiwen Ding\IEEEauthorrefmark{1},
Rajeev Thakur\IEEEauthorrefmark{2},
Le Chen\IEEEauthorrefmark{2}}
\IEEEauthorblockA{\IEEEauthorrefmark{1}\textit{University of Minnesota, Twin Cities}, Minneapolis, MN, USA}
\IEEEauthorblockA{\IEEEauthorrefmark{2}\textit{Argonne National Laboratory}, Lemont, IL, USA}
\IEEEauthorblockA{\IEEEauthorrefmark{3}\textit{The University of Chicago}, Chicago, IL, USA}
\IEEEauthorblockA{\{luo00466, dingc\}@umn.edu, \{elihu, venkat, thakur, lechen\}@anl.gov}
}

\maketitle

\begin{abstract}
Porting GPU kernels across architectures requires architectural remapping, not
syntax substitution. CUDA encodes decomposition, locality, and synchronization
through threads, blocks, and memory accesses; the Cerebras Software Language
(CSL) requires explicit placement, distributed SRAM, fabric communication,
event-driven tasks, and host/device contracts. We present \xbench, 49 paired
CUDA-to-CSL tasks, and \xkernel, an agentic framework combining target
knowledge, execution, failure-directed repair, and correctness-gated
optimization. On a fixed 28-task Level~1--3 core comparison with Claude Opus 4.8, \xkernel
raises success from 6/28 to 26/28; 22 successful programs match or beat their
CSL references. Across the 49-task coverage evaluation, 38 tasks produce a
correct program; the final three tasks are evaluated over three seeds and pass
8/9 runs. For 27
generated/reference pairs with device-internal timing, geometric-mean speedup
is 3.75$\times$ on the SDK simulator and 3.47$\times$ on WSE-3 hardware. With
the executable workflow fixed, Claude Opus~4.8 passes 26/28 core tasks while
the best open-weight model passes 2/28; retrieved Cerebras knowledge separately
raises success from 1/15 to 7/15 on a Level~1--3 panel. These results identify
base-model capability, target knowledge, execution feedback, and
same-target measurement as central to cross-architecture kernel generation.
\end{abstract}

\begin{IEEEkeywords}
agentic systems, code translation, CUDA, Cerebras Software Language,
wafer-scale computing, kernel optimization
\end{IEEEkeywords}

\section{Introduction}
Recently, due to the rapid development of Artificial Intelligence (AI), the scaling up of Machine Learning (ML) applications, and AI infrastructure, AI-infra has been recognized as the pivotal backbone of the industry. GPUs such as Nvidia's draw researchers' interest, with CUDA programming \cite{nvidiaCudaGuide} as a key initiative to drive AI-infra forward. CUDA programs reflect Nvidia GPUs' architecture through grids,
blocks, threads, memory spaces, etc., framing a hierarchical, sophisticated system task to develop high-performance GPU kernels. As a potential alternative that specializes in inference service end,
accelerator programs supported by their custom system languages present different challenges in both algorithm design and cross-architecture mapping. Porting CUDA kernels to Cerebras Software Language (CSL) \cite{cerebrassoftware,cerebrasDebug} programs is therefore a hardware/software co-design task.

Unlike Nvidia GPUs, the Cerebras Wafer-Scale Engine (WSE) \cite{cerebrasArchitecture}, a very different class of AI accelerator, places its kernels on a
physical PE mesh with distributed SRAM, fabric routes, and
event-activated tasks. A complete CSL kernel bundle couples layout parameters,
data output, input, and its movement. Translation
therefore turns virtual GPU work into explicit spatial data accesses and dataflow largely customized for inference of large language models (LLMs).

Regardless of their relatively low cost and accessibility for large data center operations, CSL programming resources are particularly scarce compared to CUDA, which already enjoys an enormous developer community and abundant codebases to learn and implement. That imbalance has given rise to emerging coding agent systems \cite{cudaforge,qimeng,stitchcuda}, and their corresponding benchmarks \cite{kernelbench}, and it is also why those systems do not yet reach a machine like the WSE.

\begin{figure*}[tb]
  \centering
  \includegraphics[width=0.98\textwidth]{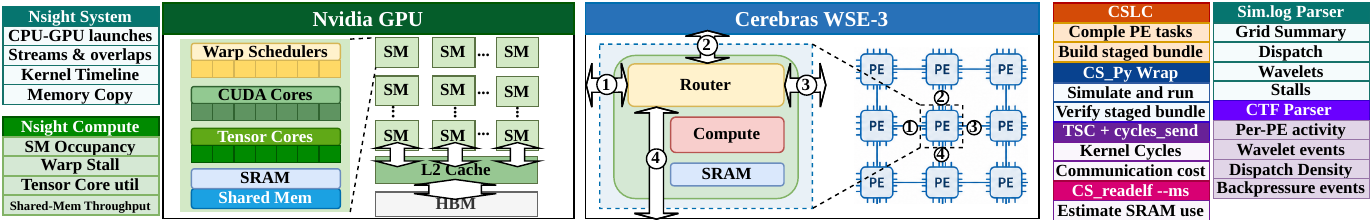}
  \caption{CUDA and Cerebras expose different execution and observation
  models. Translation reconstructs source intent and synthesizes target
  placement, storage, communication, and tasks; optimization consumes
  architecture-specific evidence rather than applying a syntax dictionary.}
  \label{fig:toolchain}
\end{figure*}

Three obstacles follow, and they organize this paper. \textbf{First, there is
almost nothing to learn CSL from.} The public corpus is essentially the official
Cerebras SDK examples and tutorials plus a handful of third-party skill notes,
with no community codebase behind it; a single generation attempt from a
frontier model therefore succeeds on only 6 of our 28 Level~1--3 core tasks. \textbf{Second,
today's coding agents were built for CUDA and carry very little CSL knowledge
into the task.} In our runs, they recover local arithmetic and individual CSL
calls but cannot assemble parameters, tasks, queues, colors, collectives,
exports, and transfers into one program that runs: across eleven models, Claude
configurations pass 24--26 of the 28 core tasks, open-weight models pass at most
2 on that same set; on a separate 12-task probe, one open model passes one task
and the other five pass none. Both sets span Levels~1--3. Moreover,
130 of 189 classified repairs are local CSL
implementation fixes rather than algorithmic ones. \textbf{Third, no benchmark
measures any of this.} KernelBench \cite{kernelbench} supplies workload semantics with no
wafer-scale target, and the SDK examples supply native spatial mechanisms with
no paired CUDA input, so there is no way to tell whether an agent has actually
learned to map one architecture onto the other. Fig.~\ref{fig:toolchain} shows
why that mapping is so hard to learn: the two sides differ not only in how a
kernel runs but in what a developer can even measure, so the tools and the
performance vocabulary an agent absorbed from CUDA have no counterpart on the
WSE.

We therefore present \xbench and \xkernel, each part answering one obstacle.
\xbench contains 49 paired CUDA and CSL tasks, each with a reference
implementation, a task description, and uniform, fair metrics for measuring how
well LLMs and agentic frameworks perform; Fig.~\ref{fig:coverage} places it
against KernelBench and the Cerebras SDK examples, which each supply only half
of what such a benchmark needs. \xkernel, our CUDA-to-CSL translation and
optimization framework, is shown end to end in Fig.~\ref{fig:overview}: it
carries a task all the way to a finished kernel, first translating the CUDA and
then optimizing the CSL against both the fabric simulator and a real WSE
cluster. It gives substantial new capability to frontier and open-source models
alike, and leaves abundant room for post-training on the very large volume of
execution traces it generates along the way.

This paper makes three contributions:
\begin{itemize}\setlength{\itemsep}{0pt}\setlength{\parskip}{0pt}
  \item \textbf{A benchmark for cross-architecture translation.} \xbench pairs
  49 CUDA kernels with validated CSL implementations across multiple operation
  families and WSE execution structures, scoring each by compiling, running, and numerically
  checking the generated program rather than by comparing text, with held-out
  inputs and safeguards that keep a model from gaming the score.

  \item \textbf{A training-free, knowledge-supported agentic framework.}
  \xkernel retrieves from a consolidated knowledge base of SDK documentation,
  tutorials, skill notes, and experience gathered from its own earlier
  knowledge-free attempts. With Sonnet 4.6, the reviewer, and a 20-attempt
  budget fixed, retrieved Cerebras material raises success from 1/15 to 7/15
  on a communication-weighted Level~1--3 panel; a
  longer repair loop alone does not recover the missing target idioms. Correct
  programs then enter a cycle-gated optimization loop
  (Fig.~\ref{fig:overview}).

  \item \textbf{An evaluation on real Cerebras tooling.} Correctness rises from
  6/28 to 26/28 on the fixed Level~1--3 core set. Across the 49-task coverage
  evaluation, 38 tasks yield a correct program, and the final three tasks pass
  8/9 seed runs. For 27
  timing-comparable generated/reference pairs, the geometric-mean speedup is
  3.75$\times$ on the simulator and 3.47$\times$ on WSE-3 hardware
  (Fig.~\ref{fig:hardware-all}). In the preliminary cross-model study, the
  three Claude models pass 24--26/28 core tasks, compared with 1--2/28 for the
  two open-weight models evaluated on the same set.
\end{itemize}

\section{Background and Related Work}
\label{sec:architecture}

\subsection{GPU and WSE Execution Models}

CUDA, the most widely used GPU programming language, exposes a virtual
grid--block--thread hierarchy: threads execute in 32-thread warps, blocks are
assigned to streaming multiprocessors at run time, and recent devices add
thread-block clusters, distributed shared memory, and asynchronous tensor
transfers. Occupancy, memory coalescing, cache reuse, divergence, and
synchronization describe the performance
tradeoffs~\cite{nvidiaCudaGuide,nvidiaBlackwellGuide}.

The WSE-3, on the other hand, integrates roughly 900,000 compute cores and 44~GB of distributed
SRAM~\cite{cerebrasWSE3}, arranged as a two-dimensional mesh of independent
processing elements (PEs) with local SRAM and statically configured fabric
routes~\cite{cerebrasArchitecture}. CSL exposes this spatial machine directly:
a layout program assigns code to PE coordinates and configures routes, while PE
programs hold local buffers, use data structure descriptors for bulk memory
operations, and bind tasks to queues or events. Host-visible symbols and
entry points complete the program~\cite{cerebrasSdkExamples}.

Fig.~\ref{fig:toolchain} summarizes the transition: per-thread implementation becomes
persistent PE-tile execution; GPU memory become PE-local SRAM and
descriptor-addressed buffers; and conditions and event triggers become colors, queues,
wavelets, routes, etc. Dynamic, parallel scheduling's focus becomes compile-time data routing and movement along with event-driven execution.

The right side of Fig.~\ref{fig:toolchain} carries the consequence that matters
most for an automated translator: the two machines are not only executed
differently, they are observed differently. A CUDA developer reasons with
Nsight statistics: bandwidth occupancy, memory throughput, and warp-level parallelism; none of those components
exists on the WSE, where the profiler provides compiler diagnostics, fabric-simulator
cycle counts, timestamp counters, wavelet and instruction traces, and ELF memory
metadata. The two optimization vocabularies barely overlap---``improve occupancy''
has no direct meaning to CSL, ``chain descriptor offsets'' has no sense in CUDA---so
performance knowledge learned on GPUs does not carry over, and any system that
ports kernels between the two architectures needs to collect the targets' own insights.
\begin{figure*}[t]
  \centering
  \includegraphics[width=0.98\textwidth]{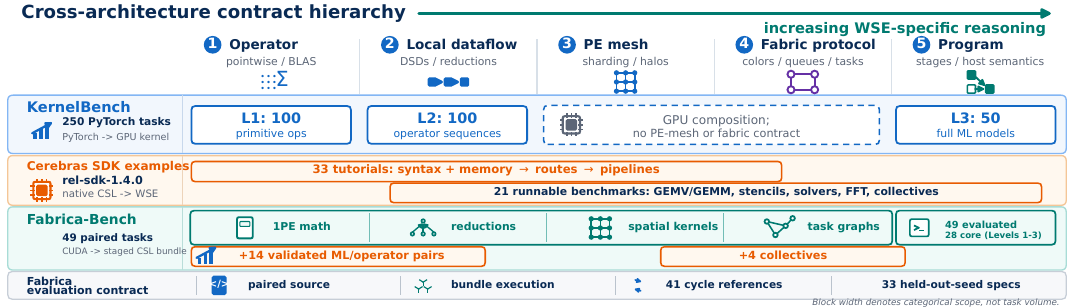}
  \caption{Where \xbench sits relative to KernelBench and the Cerebras SDK
  examples. KernelBench provides workload semantics but no wafer-scale target;
  the SDK examples supply spatial mechanisms but no paired CUDA input. \xbench
  joins the two with 49 executable pairs.}
  \label{fig:coverage}
\end{figure*}

\subsection{What Translation Needs to Preserve, Discard, and Build}

The translator needs to \textbf{preserve} cross-architecture insights such as dependencies and
I/O shape; \textbf{discard} exclusive choices of design such as a block size for WSE; and build from per-PE computation, fabric layout, to host-layout interface.
Two of these obligations have no counterpart in the source. 

Task decomposition, furthermore,
need to be re-derived rather than mapped: a CUDA block size means nothing on a
fixed PE mesh, and a thread index becomes a PE coordinate plus a local loop, so
the translator needs to recover what the kernel computes and then choose a
decomposition the source never contained. Communication needs to be invented
outright: with no shared global memory and no barrier primitive, every cross-PE
transfer becomes an explicit wavelet on a statically routed color and queue, and
a thread-block barrier becomes a routing pattern plus a task activation. With
the observability gap above, this is why translation demands working knowledge
of both architectures at once, and why a model fluent in only one of them
produces CSL that looks plausible and does not run.
Table~\ref{tab:mapping-cases} makes the gap concrete on four kernels used
throughout this paper; in every case the following rows are something the
translator needs to decide between two architectures.

\begin{table}[t]
\caption{What each CUDA kernel guarantees, and the structure the WSE program
needs to build to preserve it.}
\label{tab:mapping-cases}
\centering
\footnotesize
\setlength{\tabcolsep}{4pt}
\renewcommand{\arraystretch}{1.05}
\begin{tabularx}{\columnwidth}{@{}l@{\hspace{6pt}}Y@{}}
\toprule
\multicolumn{2}{@{}l}{\textbf{GEMV}}\\
\textsc{cuda} & Row ownership; dot-product reduction order\\
\textsc{wse}  & $8\times8$ row tiles; strided descriptors; row reduction\\
\addlinespace[3pt]
\multicolumn{2}{@{}l}{\textbf{Laplacian2D}}\\
\textsc{cuda} & Five-point dependencies; boundary semantics\\
\textsc{wse}  & Persistent PE tiles; four halo routes; receive tasks\\
\addlinespace[3pt]
\multicolumn{2}{@{}l}{\textbf{Row-parallel softmax}}\\
\textsc{cuda} & Max, subtract, exponentiate, sum, normalize\\
\textsc{wse}  & Row tiles; two all-reduces; four colors and queues\\
\addlinespace[3pt]
\multicolumn{2}{@{}l}{\textbf{Cholesky}}\\
\textsc{cuda} & Blocked factorization and update dependencies\\
\textsc{wse}  & Editable layout; both panel operands delivered to interior PEs\\
\bottomrule
\end{tabularx}
\end{table}


\xkernel records this reasoning in a source analysis $G$ (e.g.,
memory, synchronization, and GPU idioms) and the corresponding WSE design $D$ (e.g., mesh, communication, tasks, and DSD strategy). They make the mapping
inspectable but are not compiler IR: no lowering pass emits CSL.
Section~\ref{sec:mechanisms} evaluates that distinction.

\subsection{Related Work}
\label{sec:related}

KernelBench established execution-based scoring for CUDA
generation~\cite{kernelbench}, and recent agents combine staged reasoning,
profiling, and repeated execution~\cite{cudaforge,stitchcuda,kevin}. These
systems optimize inside the familiar GPU execution model. KernelCraft is closer
to our setting because it evaluates close-to-metal generation for emerging
accelerators~\cite{kernelcraft}, but starts from a specification rather than an
existing CUDA program and does not target a wafer-scale mesh.

\begin{figure*}[t]
  \centering
  \includegraphics[width=0.99\textwidth]{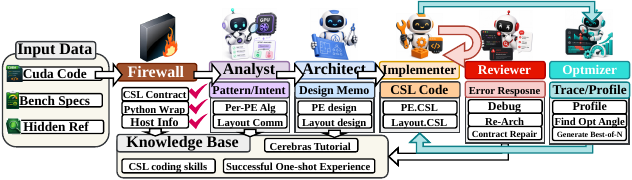}
  \caption{The \xkernel framework. A firewall removes the reference CSL and
  held-out inputs; the analyst, architect, and implementer build the CSL
  program; the reviewer classifies each failure and routes it back to the
  responsible stage; and the optimizer reduces cycles only after the program is
  correct. The knowledge base supplies the CSL knowledge the models lack.}
  \label{fig:overview}
\end{figure*}

Cross-model translation can use compiler-guided repair~\cite{unipar} or a
hand-built lowering path such as MLIR~\cite{mlir}. WSE software remains more
domain-specific: prior work implements wafer-scale stencils~\cite{rockiStencil},
models collective communication~\cite{waferScaleReduce}, lowers stencil IR to
CSL~\cite{wseMlirStencil}, or builds an inference system by
hand~\cite{waferllm}. A compiler lowering can provide stronger construction
guarantees but must be extended for each supported representation and domain.
\xkernel instead translates existing CUDA across workloads, then relies on the
target compiler, simulator, numerical tests, and hardware timing to reject
incorrect or slower candidates.

\subsection{Task Definitions and Acceptance}

The benchmark separates translation, co-design, and optimization by controlling
visibility and editable scope. In \textbf{Workflow 1 (W1)}, the fixed-layout translation, the
validated compute CSL remains hidden and only the compute file may change.
\textbf{Workflow 1-Compute (W1-C)}, the compute/layout co-design, additionally hides and makes editable
the layout, allowing placement and routing to change. In \textbf{Workflow 2 (W2)}
, the optimization from correct CSL kernels, a known-correct program is the explicit input
and the task is to reduce device cycles. This separation
distinguishes synthesis failures from optimization failures and tests whether
an apparently algorithmic failure is caused by the faulty layout.

Two terms recur below. Every candidate is staged: copied into a fresh
clone of the reference bundle and built and run by that bundle's own unmodified
scripts, so the agent cannot alter how it is tested. The \emph{host--device
interface} is what the candidate may not break: the host driver, launch order,
exported symbols, and layout-owned identifiers. Let $P$ denote a staged
candidate, $H$ its interface, and $\mathcal{I}_S$ the visible and held-out
inputs for task $S$. The primary validity gate is

\begin{equation}
\label{eq:valid}
\begin{aligned}
\pred{Valid}(P,S)\equiv{}
&\pred{Build}(P)\wedge\pred{Run}(P)\wedge\pred{Contract}(P,H)\\
&{}\wedge\pred{Timing}(P,S)\wedge\bigwedge\nolimits_{u\in\mathcal{I}_S}\pred{Correct}(P,u).
\end{aligned}
\end{equation}
where the predicates require successful compilation and execution, preservation
of that interface, a valid \emph{timing path}---the edited code region the
cycle counter actually measures---and numerical agreement on visible and
held-out inputs. For correctness-only tasks,
$\operatorname{Timing}$ checks the success gate but does not require a cycle
value. A candidate is never scored for speed unless
Eq.~\ref{eq:valid} holds.

Optimization uses the same gate. If $c_j(P)$ is the device-cycle count from
repeat $j$, a proposed edit $Q$ replaces the current best $P$ only when
\begin{equation}
\label{eq:accept}
\begin{aligned}
\operatorname{Accept}(Q\mid P,S)\equiv{}&\operatorname{Valid}(Q,S)\\
&{}\wedge \underset{j}{\operatorname{med}}\,c_j(Q)
<\underset{j}{\operatorname{med}}\,c_j(P).
\end{aligned}
\end{equation}
The inner search may rank candidates without repeating every held-out input,
but a winning edit needs to pass the held-out gate before acceptance. Thus an
invalid fast path cannot become the optimizer state.

\section{Fabrica-Bench and the Fabrica Agent}
\label{sec:system}

\subsection{Fabrica-Bench Construction}

\xbench contains 49 executable CUDA/CSL pairs, all now included in the coverage
evaluation; 28 form the fixed Level~1--3 core study used for paired workflow
and model comparisons
(Fig.~\ref{fig:coverage}). Its 31 original pairs draw mainly from SDK examples
and scientific kernels; 18 additions cover machine-learning operators and
collectives. The additions were model-assisted but manually integrated and
validated end to end, and are not treated as independent expert references.
Each task carries
its CUDA source, a validated CSL bundle, and benchmark metadata recording shapes,
split, reference cycles, held-out seeds, and the ordinal construction level
in Table~\ref{tab:construction-levels}. This suite-specific label summarizes
required CSL/WSE structure, independently of expected speedup.

\begin{table}[tb]
  \caption{Construction levels. C/D/K/F give distinct-task counts for the
  core, diagnostic, knowledge, and full sets; D has five runs per task.}
  \label{tab:construction-levels}
  \centering
  \footnotesize
  \setlength{\tabcolsep}{3pt}
  \renewcommand{\arraystretch}{0.92}
  \begin{tabularx}{\columnwidth}{@{}c>{\RaggedRight\arraybackslash}X
      >{\RaggedLeft\arraybackslash}p{0.24\columnwidth}@{}}
    \toprule
    Level & Required structure (example) & C/D/K/F \\
    \midrule
    1 & PE-local loop/reduction (ReLU) & 5/4/3/11 \\
    2 & Local dataflow/fabric stage (GEMV-row) & 20/3/10/29 \\
    3 & Custom fabric/pipeline (PDFT) & 3/1/2/8 \\
    4 & Dependent collectives (RowParallel-Softmax) & 0/0/0/1 \\
    \midrule
    Total & & 28/8/15/49 \\
    \bottomrule
  \end{tabularx}
\end{table}

The frozen core set keeps paired comparisons identical. Five repeats of the
eight-task diagnostic set expose generation variance; 12/15 knowledge tasks
are Levels~2--3, where target idioms matter most. Only the full set covers all
49 executable tasks and all four levels.

The base score,
\fastzero, requires that a program compile, run, and match the reference
numerically; performance uses in-kernel cycles rather than simulator wall time,
and higher scores require reference parity or speedup.

Six integrity checks detect reference leakage, hide evaluation inputs, freeze
timing and interface functions, reject implausible disabled timers, verify
numerical outputs, and reject runners that report success without checking them.
\begin{figure*}[tb]
  \centering
  \includegraphics[width=0.99\textwidth]{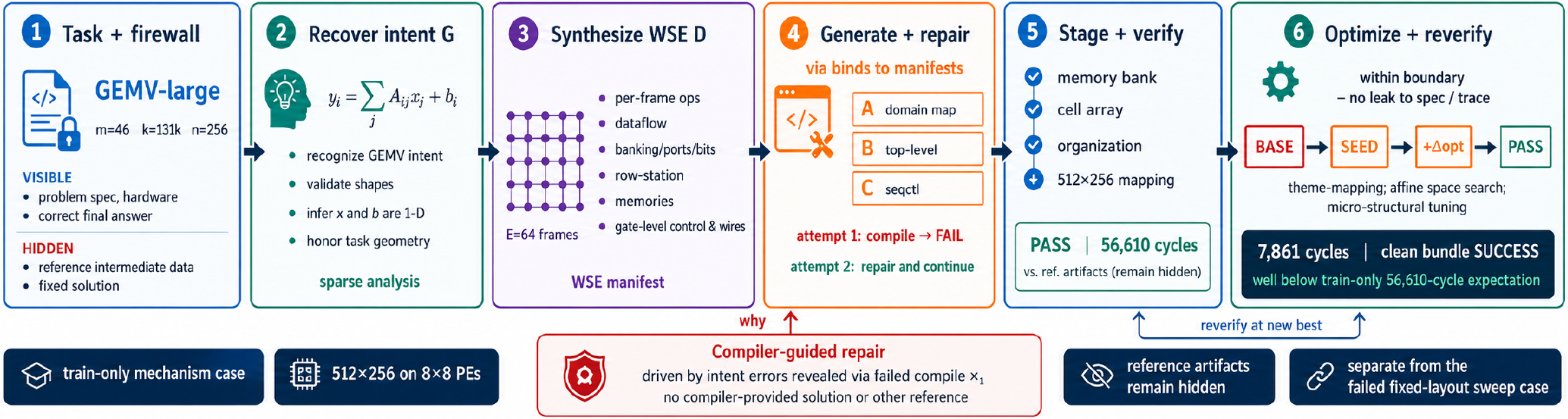}
  \caption{A large GEMV translated by \xkernel: recover a row-wise reduction,
  build an $8\times8$ collective decomposition, repair a compiler-reported API
  error, verify at 58,610 cycles, then reach 7,861 cycles in three
  correctness-checked edits against a 10,309-cycle expert reference.}
  \label{fig:workflow}
\end{figure*}

\subsection{System Overview}

Fig.~\ref{fig:overview} shows the whole framework. A task enters as a CUDA source,
its specification, and the permitted host--device interface, and a
reference-isolation firewall withholds the validated compute CSL and the
held-out inputs before anything reaches a model. Six specialized stages then run
in a fixed order: \textcircled{1} the analyst recovers the per-PE algorithm and \textcircled{2} the multi-PE
communication the source implies, \textcircled{3} the architect turns that into a placement and
layout design, \textcircled{4}the implementer writes the PE and layout CSL, \textcircled{5} the reviewer reads
compiler and runtime output and decides which stage acts next, \textcircled{6} and the optimizer
proposes cycle-reducing edits once a program is already correct. A knowledge base of CSL coding skills, Cerebras tutorials, and previously successful, preliminary knowledge-free translations feeds the stages that need it, and the reviewer's findings feed back into it.

One orchestrator invokes six role-specific model calls rather than six
persistent agents, following role specialization in multi-agent software
systems~\cite{metagpt}. Deterministic boundaries enforce reference isolation,
route execution failures, and accept an optimization only when it remains valid
and reduces median cycles (Eq.~\ref{eq:accept}).

\subsection{A Task Through Fabrica: GEMV-Large}

Fig.~\ref{fig:workflow} follows a $512\times256$ GEMV through the six stages.
The analyst preserves the independent output rows and the reduction over matrix
columns; the architect tiles the matrix across $8\times8$ PEs, broadcasts the
vector and bias, reduces by row, and gathers from the last column; compiler
feedback then repairs a mis-named parameter before the original runner verifies the result. The figure's cycle counts show what each phase contributes: 58,610 cycles at first correctness, 7,861 after three optimization edits, against a
10,309-cycle expert reference. This train-only trace is distinct from the numerically invalid fixed-layout GEMV in the coverage sweep.

The workflow separates transient state, reusable knowledge, and verify history.
Per-run memory holds $G$, $D$, the current candidate, the latest failure, and
bounded profile evidence. Because CSL is largely absent from pretraining data,
the knowledge base needs to carry the language itself: SDK documentation and release
notes, published tutorials and examples, and third-party CSL skill notes.
Retrieving documentation to write code for unfamiliar APIs follows
DocPrompting~\cite{docprompting}; here it is the primary remedy for a target the
model has essentially never seen. Retrieval is filtered by role, so the
architect receives decomposition patterns, the implementer CSL and API rules,
and the reviewer failure recipes. Worked examples are leave-one-out and drawn
only from the training pool, and complete reference bundles never enter
retrieval. An append-only storage records prompts, candidates, outcomes, and
cycles, but prior code returns only as screened aggregate lessons, adapting
reflective memory~\cite{reflexion} to a setting where an earlier program may be
the answer to the hidden benchmark.

\subsection{Constrained Generation and Execution Feedback}

The implementer synthesizes a complete compute file. It may optionally begin
from one of three starter templates---single-PE computation, nearest-neighbor
halo exchange, or a two-dimensional collective---each supplying common imports,
timing hooks, task declarations, and host-interface structure with
kernel-specific slots left blank. The model actor still returns a complete file,
so repair can overwrite the supplied structure. Across the 49-task evaluation,
47 tasks receive the generic single-PE template, one starts from scratch, and
one lacks a selector record. We therefore analyze this treatment as a stable
starting structure followed by repair rather than as evidence for the
specialized templates.

\begin{table*}[tb]
\caption{Experimental protocol. Parenthesized task counts are ordered by
construction level, with trailing zeros omitted. Each row keeps its own
denominator. T/O is the translation/optimization attempt cap.}
\label{tab:protocol}
\centering
\footnotesize
\setlength{\tabcolsep}{4pt}
\renewcommand{\arraystretch}{1.15}
\begin{tabularx}{\textwidth}{@{}L{0.84}L{1.00}L{0.90}L{0.82}L{1.44}@{}}
\toprule
\textbf{Study} & \textbf{Benchmark set} & \textbf{Workflow entries} &
\textbf{Budget (T/O)} & \textbf{Reported outcome} \\
\midrule
\grouphead{End-to-end translation}
Controlled comparison & Core 28 (5/20/3) & 28 single-attempt; 28 full & 1/0; 40/40 & Bundle pass; faster-than-reference \\
Model study           & Core 28 (5 models); probe 12 (6 models; 1/9/2) & 212 model--task & 40/40 & Bundle pass; faster count per model \\
\addlinespace
\grouphead{Ablations}
Feedback ablation  & Diagnostic 8 (4/3/1) & 40 per mode (5/task; A--D, T) & A/C 4/0; B/D/T 10/10 & Retained bundle pass \\
Knowledge ablation & Knowledge 15 (3/10/2) & 45 translations       & 20/20                & Pass rate per knowledge tier \\
\addlinespace
\grouphead{Held-out coverage}
Coverage evaluation & Full 49 (11/29/8/1) & 55 workflows (3 tasks $\times$ 3 seeds) & 10/10 & 38 tasks with a pass; 8/9 seeded runs \\
\addlinespace
\grouphead{Optimization and supporting studies}
Optimization only    & 14 references (2/12) & One optimization each & 0/10 & Accepted speedups (correctness-safe) \\
Hardware replay      & 35 generated/ref. pairs  & 140 bundle--target runs & Replay only     & 27 timing-comparable speedup pairs \\
Offline SFT pilot    & 6 held-out tasks; 8B/14B & 96 outputs (48 pairs)   & 4 per condition & SDK-free CSL proxy metrics \\
\bottomrule
\end{tabularx}
\end{table*}

\begin{table}[t]
\caption{Complete-system and component studies. Denominators 28 and 15 count
tasks; 40 counts five runs on each of eight diagnostic tasks.}
\label{tab:headline}
\centering
\footnotesize
\setlength{\tabcolsep}{3pt}
\begin{tabular}{@{}lrrl@{}}
\toprule
Comparison & Base & Changed & Gate \\
\midrule
Complete workflow & 6/28 & 26/28 & final \\
Retrieved knowledge & 1/15 & 7/15 & final \\
Design records, unguided & 21/40 & 20/40 & final \\
Design records, guided & 33/40 & 32/40 & translate \\
Starting-template mode & -- & 30/40 & final \\
\bottomrule
\end{tabular}

\end{table}

Before generation, a deterministic extractor scans only visible files for
compile-time parameters, host launch order, exported symbols, layout-owned task
and color identifiers, and reserved queues. A second option parses $D$ into
mesh, shard, buffer, collective, task, color, and queue records, then checks
mesh agreement, the 48-KB per-PE SRAM budget, and resource collisions. These
records are prompt guidance, not compiler IR: parsing is best-effort and no
lowering pass constructs CSL, so the feedback ablation in
Section~\ref{sec:mechanisms} measures whether the extra structure helps the
model rather than crediting it with compiler guarantees.

Every candidate is staged in a temporary reference bundle and checked by its
original build, runner, and numerical tests; malformed CSL first fails a
compile-only gate. The reviewer then routes architectural, implementation, and
host/layout failures to redesign, code repair, or interface repair,
respectively~\cite{selfdebug}. The transcript retains the candidate, status,
held-out result, and device cycles.

Correct candidates may then be profiled. The feedback is bounded and
heuristic---event mixes, dispatch-span proxies, emitted wavelet or backpressure
signals, and SRAM estimates---rather than calibrated occupancy or congestion
counters, and the interactive debugger, visualizer, and cycle-by-cycle traces
remain aids for a human rather than automatic signals~\cite{cerebrasDebug}.

The optimizer ranks permitted transformations using a static CSL scan and,
when enabled, profile evidence. It restores frozen timing and interface state,
executes each proposal, and retains only a valid lower-median candidate;
patience bounds an unproductive search. Architecture mapping therefore precedes
implementation, correctness precedes optimization, and every accepted change
satisfies Eq.~\ref{eq:accept}.

\subsection{From Run Logs to Post-Training Data}

Every workflow leaves a complete run log: the prompts, the CSL each attempt
produced, the compiler and runtime feedback, and the outcome. The exporter
aligns each code-producing turn with its saved attempt and, for turn $a_i$ in
run log $\tau$, assigns
\begin{equation}
r_{\tau}(a_i)=
\begin{cases}
1, & \operatorname{Pass}(a_i),\\
0, & \neg\operatorname{Pass}(a_i)\ \wedge\
     \exists j>i:\operatorname{Pass}(a_j),\\
-1, & \text{otherwise}.
\end{cases}
\label{eq:trajectory-reward}
\end{equation}

This deterministic label records whether a turn is correct or belongs to a
repair sequence that eventually succeeds; it is not a learned reward model.

\begin{figure*}[t]
  \centering
  \includegraphics[width=0.99\textwidth]{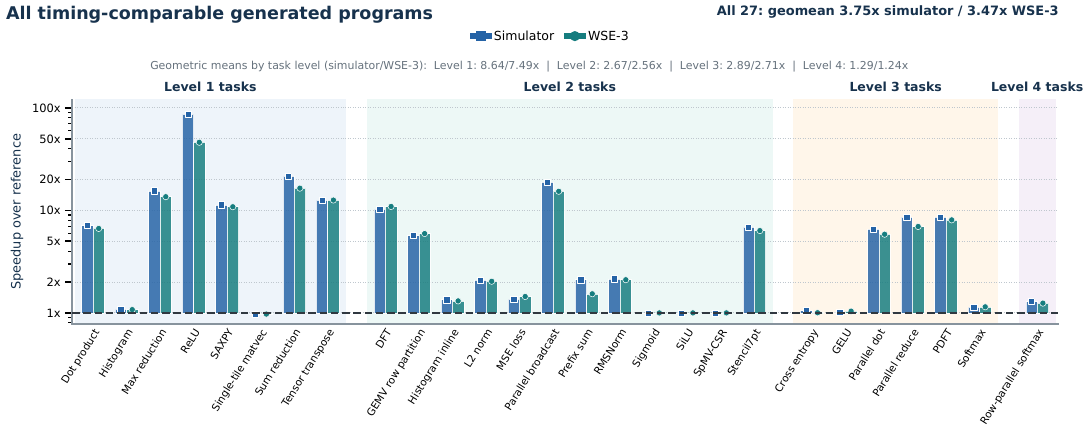}
  \caption{All 27 timing-comparable generated/reference pairs, grouped by the
  four construction levels in each task's \texttt{spec.yaml}. Every level band
  shares one logarithmic axis; paired vertical bars compare simulator and WSE-3
  speedups, and the dashed line marks parity with the reference. The levels
  describe program-construction complexity rather than expected performance
  gain.}
  \label{fig:hardware-all}
\end{figure*}

Each reduction is explicit. Of 1,741 run directories, 1,673 align; eventually
passing runs yield 7,619 supervised candidates; and removing 306 duplicates and
1,247 records above the 500-per-kernel cap leaves 6,066, of which 2,440 are
verified passes and 3,626 intermediate repairs. No turn from an all-failure run
enters the set. For preference optimization, the builder requires the same prompt
prefix and an adjacent failed-to-passing transition, choosing the passing
response and rejecting the failed one~\cite{dpo}, which after deduplication
gives 2,249 pairs. Short responses, suspected reference leakage, and
architecture-reset turns are filtered out. These are offline supervised and
preference records, not online reinforcement-learning episodes as used by recent
CUDA systems~\cite{kevin}.

\section{Experimental Methodology}
\label{sec:method}

Table~\ref{tab:protocol} fixes the task sets, repetitions, and attempt budgets;
results from different rows are not pooled. The primary controlled, coverage,
and optimization studies use Claude Opus~4.8 and SDK~1.4.0 simulation on fresh
bundles with the original numerical checks and disjoint train/test
splits~\cite{anthropicClaude}. The preliminary 11-model study changes only the
generation model and covers three commercial frontier models and eight
open-weight models: five models each receive the 28-task core and six additional
open-weight models each receive the same 12-task Level~1--3 probe. It runs one
workflow per model--task pair, so we compare models only within a shared task
set and do not pool the two denominators. GEMM and GEMV remain
train-only because retrieved WaferLLM patterns overlap those
families~\cite{waferllm}. Hardware replay recompiles the retained bundles under
SDK~2.10 and measures them on both its appliance simulator and WSE-3.

The single-attempt arm disables retrieval, review, repair, and optimization;
the full arm generates three candidates before bounded repair. In the feedback
study, A is unguided, B adds classified execution repair, C adds design records
to A, D adds them to B, and T adds a starting template to B. A retained pass
requires both \texttt{status=pass} and \texttt{success\_marker=true}; correctness
and cycle aggregates include only retained passes. Profiling uses ten paired
runs per original task and five per later-added task.

We report pass rate on each stated denominator and geometric-mean
reference-to-generated speedup over valid cycle measurements. Hardware rows use
the same SDK~2.10 bundle and kernel-owned \texttt{run.py} on both targets; all
four generated/reference simulator/hardware runs must pass with plausible
device-internal timing. Host-bracketed RPC timing is excluded. Full prompts,
candidates, checks, and measurements remain in the artifact.

\section{Evaluation and Results}


\subsection{End-to-End and Cross-Model Results}

Table~\ref{tab:headline} compares the same model on the fixed 28-task
Level~1--3 core (5/20/3 by level). A single generation attempt passes 6
programs (21\%) and produces one result faster than
its reference. The full workflow passes 26 (93\%), a 72 percentage-point
correctness difference; 22 of those 26 match or beat their reference and 16 are
strictly faster. This complete-system result captures the combined contribution
of retrieval, multiple candidates, staged repair, and optimization; the
controlled rows below examine individual mechanisms.

Table~\ref{tab:models} reports the preliminary cross-model comparison. On the
identical 28-task core, Opus~4.8 passes 26 tasks, Sonnet~4.6 passes 25, and
Opus~4.6 passes 24; the best open-weight model, gpt-oss-120b, passes 2. Thus the
best commercial-to-open comparison differs by 24 tasks, or 86 percentage points,
under the same workflow. The three Claude models are much closer to one
another, and Sonnet~4.6 produces 17 faster-than-reference programs versus 16
for Opus~4.8. On the fixed 12-task probe, gpt-oss-20b passes one task and the
other five open models pass none. This single-workflow-per-task experiment
establishes a large commercial-to-open capability gap rather than a statistical
ranking among the frontier models.

\begin{figure*}[t]
  \centering
  \includegraphics[width=0.385\textwidth]{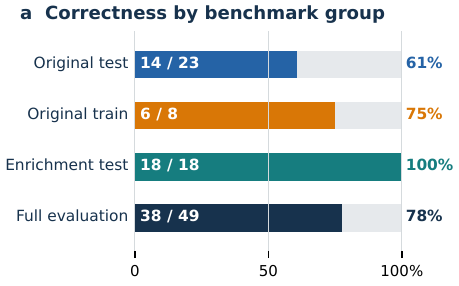}\hfill
  \includegraphics[width=0.585\textwidth]{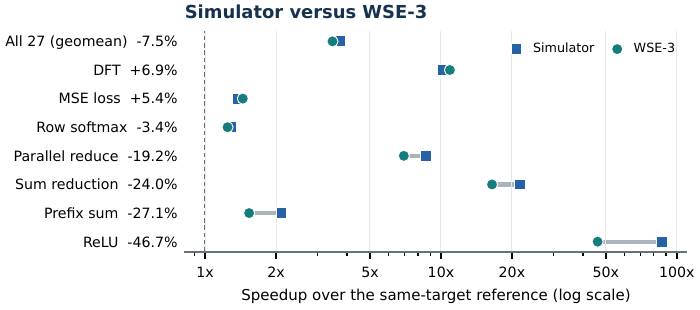}
  \caption{Complementary evaluations. Left: task-level correctness across the
  49-task coverage evaluation. Right: paired same-target speedups on the SDK~2.10
  simulator (squares) and WSE-3 (circles). The aggregate uses all 27 comparable
  pairs; selected kernels show the two largest increases, a distributed
  near-agreement case, and the four largest decreases.}
  \label{fig:templatesweep}
\end{figure*}

\begin{table}[t]
\caption{Preliminary cross-model comparison with the full executable workflow.
The core and probe are reported separately. $>$Ref counts correct programs
faster than their same-target reference.}
\label{tab:models}
\centering
\footnotesize
\setlength{\tabcolsep}{3pt}
\renewcommand{\arraystretch}{1.02}
\begin{tabularx}{\columnwidth}{@{}Xrr@{}}
\toprule
\textbf{Generation model} & \textbf{Pass} & \textbf{$>$Ref} \\
\midrule
\multicolumn{3}{@{}l}{\itshape Core 28: five models, identical Level~1--3 tasks} \\
Claude Opus 4.8      & 26/28 & 16 \\
Claude Sonnet 4.6    & 25/28 & 17 \\
Claude Opus 4.6      & 24/28 & 14 \\
gpt-oss-120b         &  2/28 &  1 \\
Gemma 4 31B          &  1/28 &  0 \\
\addlinespace
\multicolumn{3}{@{}l}{\itshape Probe 12: six models, identical Level~1--3 tasks} \\
gpt-oss-20b          &  1/12 &  0 \\
Other five open models & 0/12 each & 0 \\
\bottomrule
\end{tabularx}
\end{table}

\subsection{Target Knowledge Drives the Strongest Controlled Gain}

The knowledge ablation isolates the strongest single component. Across its
Level~1--3 panel (3/10/2) and 45 translations, with the Sonnet 4.6 model,
reviewer, and 20-attempt budget fixed, full knowledge passes 7/15; removing
retrieved Cerebras
material passes 1/15, and also removing the CSL primer remains at 1/15. The six
additional successes all require collective operations, halo exchange, or
DSD/queue setup; the knowledge-free condition succeeds on a one-PE
matrix-vector kernel. Retrieved implementation patterns therefore account for
six additional successes on communication- and dataflow-intensive programs
under the fixed repair budget.

\subsection{Feedback Changes Where Search Effort Goes}
\label{sec:mechanisms}

Modes A--D vary guided execution feedback and structured design records on the
Level~1--3 diagnostic panel (4/3/1), with five repeats per task and 40 runs per
mode. The records axis provides the
single-variable comparison: adding the records changes unguided retained
success from 21/40 to 20/40 and guided translation-stage success from 33/40 to
32/40. The records make each proposed mapping inspectable,
while measured correctness remains effectively unchanged. The starting-template
mode retains 30/40 final passes, including all 15 repeats of the three simplest
Level~1 operators and two Game-of-Life runs. The subsequent coverage evaluation applies the same
starting-program and repair mechanism across all 49 swept tasks.

The reviewer classifications partition 189 broader-sweep repair requests:
130 (68.8\%) are local implementation repairs, 37 (19.6\%) cross a bundle
boundary, and 22 (11.6\%) request a new decomposition. The compile-only gate
filters 140/224 malformed candidates before simulation. Workflows that reach a
retained pass converge in a median of two evaluations, while the remaining 11
use the full ten-evaluation budget. This distribution supports a short initial
path followed by additional repair only when staged evidence requests it.

\subsection{A Starter Template and Repair Broaden Coverage}

Across all 49 swept tasks, 38 produce at least one correct program (78\%).
Fig.~\ref{fig:templatesweep} separates 14/23 original held-out tests, 6/8
original training programs, and 18/18 later-added tasks, giving 20/31 on the
original set and 32/41 across all held-out tasks. The added references were
integrated by hand after model-assisted first translations and are reported as
a separate benchmark group.

A three-seed follow-up adds the final three tasks: GEMM-1PE passes 3/3 at a
7.64$\times$ geometric-mean speedup,
Laplacian2D-Reduce passes 3/3 at 1.78$\times$, and LorenzoPredictor-Tile passes
2/3 at 2.36$\times$. Together they pass 8/9 runs, and every passing run beats
its reference.

Of the 32 passing held-out tasks, 28 use fewer cycles than their references,
two match them, and two use more. The 47/49 generic-template selection shows
that measured coverage primarily reflects a stable initial structure followed
by execution-guided repair.

RowParallel-Softmax illustrates the repair sequence. Seven attempts expose, in
turn, a layout/PE parameter mismatch, asynchronous-task and descriptor errors,
a numerically wrong distributed reduction, and scope/type faults. Attempt eight
passes three hidden seeds at 17,892 cycles; two accepted edits reach 13,688
cycles, 1.29$\times$ faster than the 17,714-cycle reference. The sequence shows
how successive compiler, numerical, and cycle evidence completes the
distributed contract.

GEMM-Collectives-2D reaches 48,188 cycles against 51,738 (1.07$\times$);
GEMV-Collectives-2D reaches 4,336 against 4,003 (0.92$\times$), separating
correctness from speed. Game-of-Life adds 2/5 passing runs in a repeated
follow-up.

\subsection{Correctness-Gated Optimization}

When W2 begins from correct CSL, the optimizer improves all 14 measured
references, with a 1.44$\times$ geometric-mean speedup and seven programs
improving by at least 25\%. This isolates optimization headroom from
translation success. PE-local arithmetic exposes the largest opportunities:
bulk DSD operations, fused arithmetic, loop simplification, and compile-time
constants can remove substantial work. Library-delegated and
communication-dominated timing paths more often remain near parity.
Across the feedback-guided modes and coverage sweep, correctness and cycle
gates accept 110/601 proposed edits (18\%). Descriptor-offset chaining, task
simplification, and buffer cleanup are accepted at 27/97, 20/65, and 20/83;
bulk FMAC is accepted in 12 of 131 proposals. The accepted edits are WSE-native
transformations rather than CUDA optimizations carried across unchanged.

\subsection{Simulator Speedups Mostly Persist on WSE-3}

We replay the 35 passing bundles retained in the frozen hardware-replay set and
their references through the same SDK~2.10 host path. Twenty-seven pairs pass on both targets with comparable
device-internal timing; 23/27 generated programs are faster on hardware and
26/27 match or beat their references. Their geometric-mean speedup changes from
3.75$\times$ on the simulator to 3.47$\times$ on WSE-3
(Fig.~\ref{fig:hardware-all}).

The shift is a relative effect between two implementations. For target $t$,
let $S_t=c_t^{\mathrm{ref}}/c_t^{\mathrm{gen}}$. Then
\begin{equation}
\frac{S_{\mathrm{hw}}}{S_{\mathrm{sim}}}=
\frac{c_{\mathrm{hw}}^{\mathrm{ref}}/c_{\mathrm{sim}}^{\mathrm{ref}}}
     {c_{\mathrm{hw}}^{\mathrm{gen}}/c_{\mathrm{sim}}^{\mathrm{gen}}}.
\label{eq:target-shift}
\end{equation}
Generated and reference hardware/simulator geometric means are 1.030 and
0.953, yielding the aggregate 7.5\% reduction. Per-kernel movement can be
larger. DFT improves from 10.21$\times$ to 10.92$\times$ because its generated
program becomes relatively faster (0.934 hardware/simulator) while its
reference is nearly unchanged (0.998). RowParallel-Softmax stays close
(1.29$\times$ to 1.24$\times$; ratios 1.074 and 1.037). Prefix-Sum falls from
2.11$\times$ to 1.54$\times$ because the generated program slows to 1.111 while
the reference falls to 0.810. ReLU is the extreme: its 142-cycle generated
path rises to 222 cycles while the reference drops from 12,299 to 10,252,
reducing speedup from 86.6$\times$ to 46.2$\times$.

Fig.~\ref{fig:hardware-all} shows the complete distribution. Level~1 tasks span
the broadest range, from 0.97$\times$ to ReLU's 46.2$\times$ on WSE-3. Level~2
and Level~3 include collective, stencil, normalization, and pipeline programs
with geometric means of 2.56$\times$ and 2.71$\times$ on WSE-3. The Level~4
RowParallel-Softmax task remains close to its expert reference at
1.24$\times$. Thus, construction level and achieved speedup measure different
properties of a translated program.
Figure~\ref{fig:templatesweep} complements the complete distribution by placing
the 49-task correctness accounting beside the aggregate hardware replay and its
largest measured simulator-to-WSE shifts.

\subsection{Architecture Mapping Exposes System Boundaries}

The cases in Table~\ref{tab:mapping-cases} distinguish arithmetic,
communication, interface, and co-design boundaries. Laplacian2D, for example,
passes 5/5 under execution-guided repair compared with 0/5 in the unguided
condition, showing that staged feedback can assemble its halo contract.
\begin{figure}[tb]
  \centering
  \includegraphics[width=0.92\columnwidth]{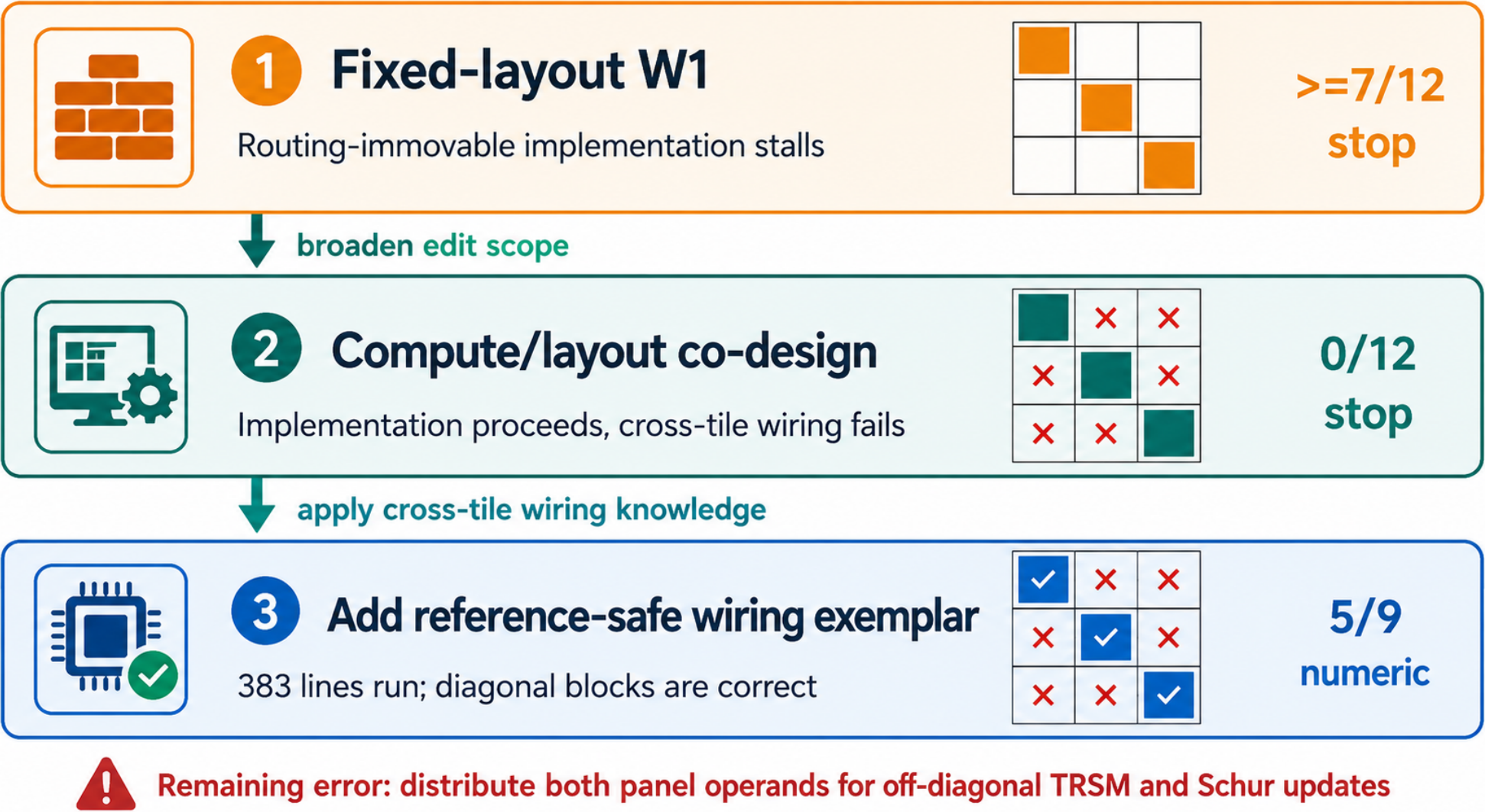}
  \caption{Cholesky co-design progression under expanding edit scope. Layout
  editing removes early interface stops; the exemplar-assisted stage produces
  a running distributed program and reaches numerical checking in 5/9 attempts.}
  \label{fig:cholesky}
\end{figure}
For Cholesky, layout editing removes at least five early interface stops
(Fig.~\ref{fig:cholesky}). Adding a train-derived
\texttt{collectives\_2d} example then produces a running 383-line co-design in
which 5/9 attempts reach numerical checking and diagonal blocks pass. The
progression separates immutable routing, cross-file wiring, and operand delivery
and identifies the final distributed-operand boundary for focused follow-up.

\subsection{Cerebras-Derived Offline SFT Pilot}

\begin{figure}[tb]
  \centering
  \includegraphics[width=0.94\columnwidth]{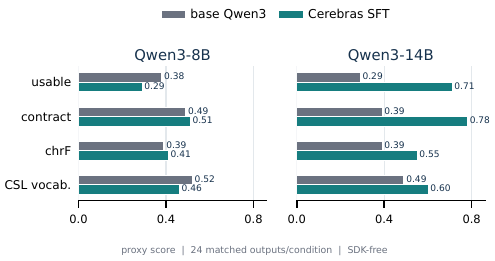}
  \caption{Cerebras-corpus SFT proxy metrics on six held-out kernels
  (24 matched outputs per condition).}
  \label{fig:sft-pilot}
\end{figure}

We derive LoRA training data~\cite{lora} from the 6,066 CUDA-to-CSL records in
Sec.~\ref{sec:system}: 2,440 verified targets and 3,626 repairs from
eventually-passing Cerebras runs. Benchmark-disjoint filtering retains 1,813 rows
from 39 kernels and excludes all six evaluation kernels. The verified 14B run
uses rank 64, $\alpha=128$, and three epochs; loss falls from 0.823 to 0.639,
and the serving log loads \texttt{checkpoint-153} rather than the base model.
For Qwen3-8B and Qwen3-14B, four generations per held-out kernel give 24
matched base/SFT outputs per model. Each pair shares its CUDA input and real CSL
reference, with fenced CSL extracted before scoring. The 14B
usable/contract/chrF/vocabulary scores rise from 0.29/0.39/0.39/0.49 to
0.71/0.78/0.55/0.60; the 8B scores change from 0.38/0.49/0.39/0.52 to
0.29/0.51/0.41/0.46 (Fig.~\ref{fig:sft-pilot}).

\section{Discussion and Limitations}
\label{sec:implications}

The results support an adaptive workflow: retrieve target knowledge, use a
short initial path for PE-local tasks, invoke classified repair after observed
compile, interface, or numerical failures, and optimize only validated
programs. 
A promising next step is to lower
typed mappings into CSL rather than append more facts to prompts, following the
direction of MLIR-based CSL compilation~\cite{mlir,wseMlirStencil}.

The 49-task coverage metric counts a
task once if any scheduled workflow passes; three tasks use three seeded
workflows and the remaining tasks use one, yielding 38/49 task successes and
8/9 seeded-run successes. 
The SFT pilot reports SDK-free structural metrics rather than executable
correctness; generated programs still require compilation, numerical checks,
and device timing.

\section{Conclusion}

\xkernel treats CUDA-to-CSL translation as distributed program construction: it
recovers source invariants, builds WSE placement and communication, rebuilds in
a sandboxed copy of the reference bundle, and optimizes only after validation.
Success rises from 6/28 to 26/28 on the fixed Level~1--3 core set, and the broader
49-task coverage evaluation yields correct programs for 38 tasks; the final
three tasks pass 8/9 seed runs. Across 27 timing-comparable pairs, the
geometric-mean speedup changes from 3.75$\times$ on the simulator to
3.47$\times$ on WSE-3 hardware. The controlled ablations identify retrieved
Cerebras knowledge as the strongest isolated factor (7/15 versus 1/15 on the
Level~1--3 knowledge panel).
The evidence favors target knowledge, short initial
generation, failure-triggered repair, gated optimization, and same-target
measurement. \xbench and its recorded runs provide a reproducible base for
cross-accelerator translation. Offline SFT improves all four 14B proxy metrics
but still requires executable CSL validation.

\section*{Acknowledgment and AI-Use Disclosure}

OpenAI Codex and Anthropic Claude assisted language editing in the Abstract and
Sections~I--VII~\cite{openaiCodex,anthropicClaude}. The authors verified all
technical claims, citations, equations, figures, tables, and results against the
source artifacts and retain full responsibility.

\bibliographystyle{IEEEtran}
\bibliography{xkernel}

\end{document}